\documentclass[prd,onecolumn,superscriptaddress,nofootinbib,12pt]{revtex4-1}
\usepackage{url}
\usepackage{amsfonts}
\usepackage{amsmath}
\usepackage{amssymb}

\usepackage{epsfig}

\usepackage{graphicx}
\usepackage{epsfig}
\usepackage{eepic}
\usepackage{color}
\usepackage{bbm}
\usepackage{dcolumn}
\usepackage{bm}
\usepackage{ulem}
\usepackage{mathrsfs}
\usepackage{bbold}
\usepackage{datetime}

\usepackage{overpic} 
\usepackage{rotating}
\usepackage[usenames,dvipsnames]{xcolor}
\usepackage[colorlinks=true,citecolor=Blue,linkcolor=Green,urlcolor=Green]{hyperref}
\usepackage{lipsum} 
\usepackage{tikz,tikz-3dplot}
\usetikzlibrary{shapes.geometric}
\usepackage{etoolbox} 
\usepackage[capitalize]{cleveref}
\usepackage{extarrows}

\def\bc{\begin{center}}

\def\ec{\end{center}}
\def\be{\begin{eqnarray}}
\def\ee{\end{eqnarray}}
\definecolor{dyellow}{rgb}{1.,0.8,.0}
\definecolor{myblue}{rgb}{.1,.1,.7}
\definecolor{dcyan}{rgb}{.0,.6,.6}
\definecolor{dmagenta}{rgb}{0.6,0.0,0.6}
\definecolor{brown}{rgb}{0.6,0.2,0.}
\definecolor{darkblue}{rgb}{.0,.0,0.5}
\definecolor{darkred}{rgb}{0.75,0.0,0.0}
\definecolor{orange}{rgb}{1.,.6,.0}
\definecolor{dorange}{rgb}{0.8,.4,.0}
\definecolor{darkgreen}{rgb}{0.0,0.6,0.0}
\definecolor{purple}{rgb}{.4,.0,.4}
\definecolor{lightgrey}{rgb}{0.7, 0.7, 0.7}
\definecolor{grey}{rgb}{0.4, 0.4, 0.4}

\usepackage{geometry}
\usepackage[position=t, singlelinecheck=off]{subfig}
\usepackage[font=small,labelfont=bf,justification=raggedright]{caption}

\newcommand{\xdownarrow}[1]{%
  {\left\downarrow\vbox to #1{}\right.\kern-\nulldelimiterspace}
}
\newcommand{\xuparrow}[1]{%
  {\left\uparrow\vbox to #1{}\right.\kern-\nulldelimiterspace}
}

\begin{document}
\newsavebox{\lefttempbox}
\title{
From black hole to one-dimensional chain: parity symmetry breaking and kink formation}
\author{Zhi-Hong \surname{Li}} \email{lizhihong@buaa.edu.cn}
\affiliation{Department of Physics, Shanxi Datong University, Datong 037009, China}
\affiliation{Institute of Theoretical Physics, Shanxi Datong University, Datong, 037009, China}
\author{Han-Qing Shi} \email{by2030104@buaa.edu.cn (Corresponding author)}
\affiliation{Center for Gravitational Physics, Department of Space Science, Beihang University,
Beijing 100191, China}
\author{Hai-Qing Zhang} \email{hqzhang@buaa.edu.cn (Corresponding author)}
\affiliation{Center for Gravitational Physics, Department of Space Science, Beihang University,
Beijing 100191, China}
\affiliation{Peng Huanwu Collaborative Center for Research and Education, Beihang University, Beijing 100191, China}

\begin{abstract}
{\centering {\Large Abstract}\\}
AdS/CFT correspondence is a ``first-principle" tool to study the strongly coupled many-body systems. While it has been extensively applied to investigate the continuous symmetry breaking dynamics, the dynamics of discrete symmetry breaking are rarely investigated. In this paper, the model of kink formation in a strongly coupled one-dimensional chain is realized from the AdS/CFT correspondence. In doing so, we first construct a model of real scalar fields with parity symmetries in the AdS bulk. By quenching the system across the critical point at a finite rate, kink hairs turn out in the bulk due to the spontaneous parity symmetry breaking, which accomplishes a counter-example of ``no hair conjecture" of black hole. Due to the AdS/CFT correspondence, kink hairs in the bulk are dual to the kinks in the AdS boundary. The mean of the dual kink numbers are found to satisfy a universal power-law relation to the quench rate, in agreement with the celebrated Kibble-Zurek mechanism. Moreover, the higher cumulants of the kink numbers are proportional to the mean numbers, consistent with the assumption that the formation of kinks satisfy the binomial distributions which goes beyond the Kibble-Zurek mechanism. 
\end{abstract}

\maketitle


\newpage

\section{Introduction}
{\it No hair conjecture} of black hole states that the solution of black hole can be entirely characterized by its mass, electric charge and angular momentum \cite{Israel:1967wq, Israel:1967za,Carter:1971zc}. However, this conjecture was challenged in various ways. Counter-examples are found in black holes with non-Abelian Yang-Mills fields \cite{Volkov:1989fi}, in higher dimensional gravity \cite{Torii:1996yi}, or with the soft hairs \cite{Hawking:2016msc}, just to name some examples. For reviews one can refer to \cite{Liu:2022eri} and references therein. Among these, Gubser's proposal \cite{Gubser:2008px} for the hair of charged scalar fields in a spacetime with negative cosmological constant was widely applied in the holographic superconducting phenomena \cite{Hartnoll:2008vx}, which brought in a surge of the applied holography especially in condensed matter physics \cite{Zaanen:2015oix}. 

However, the study of the hair of topological structures is rare \cite{Luckock:1986tr}.  In this paper, we initiate the investigation on the hair of topological defects -- kink formations from the discrete symmetry breaking in a planar Schwarzschild-Anti de-Sitter (AdS) black hole due to the prominent Kibble-Zurek mechanism (KZM) \cite{Kibble:1976sj,Kibble:1980mv,Zurek:1985qw}. KZM was first proposed in the cosmology scenario \cite{Kibble:1976sj} and later was extended to the superfluid in condensed matter \cite{Zurek:1985qw}. It states that when the system is driven into a symmetry breaking phase in a finite rate, topological defects will emerge at the interfaces between the distinct symmetry breaking domains. Therefore, the number of topological defects can be predicted theoretically which obeys a universal power-law scaling to the quench rate. KZM has been tested in various systems in condensed matter physics, such as \cite{Chuang:1991zz,Ruutu:1995qz,Carmi:2000zz}. For reviews in KZM, please refer to \cite{zurekreview,delCampo:2013nla}.

\begin{figure}[h]
\centering
\includegraphics[trim=0cm 0cm 0cm 0cm, clip=true, scale=0.7, angle=0]{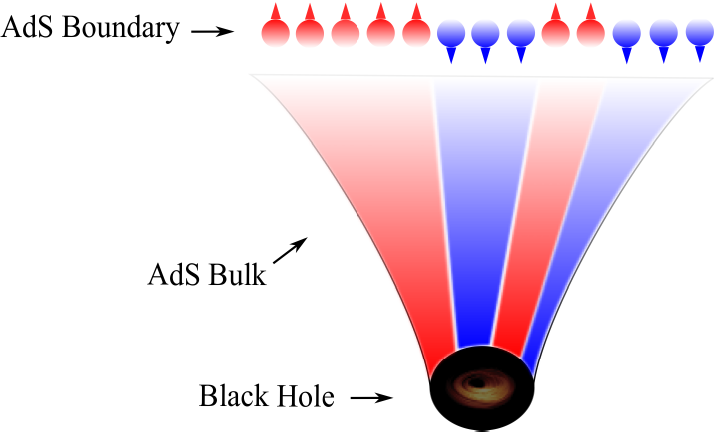}
\hspace{-1.3cm}\includegraphics[trim=0cm 0cm 0cm 0cm, clip=true, scale=0.25, angle=0]{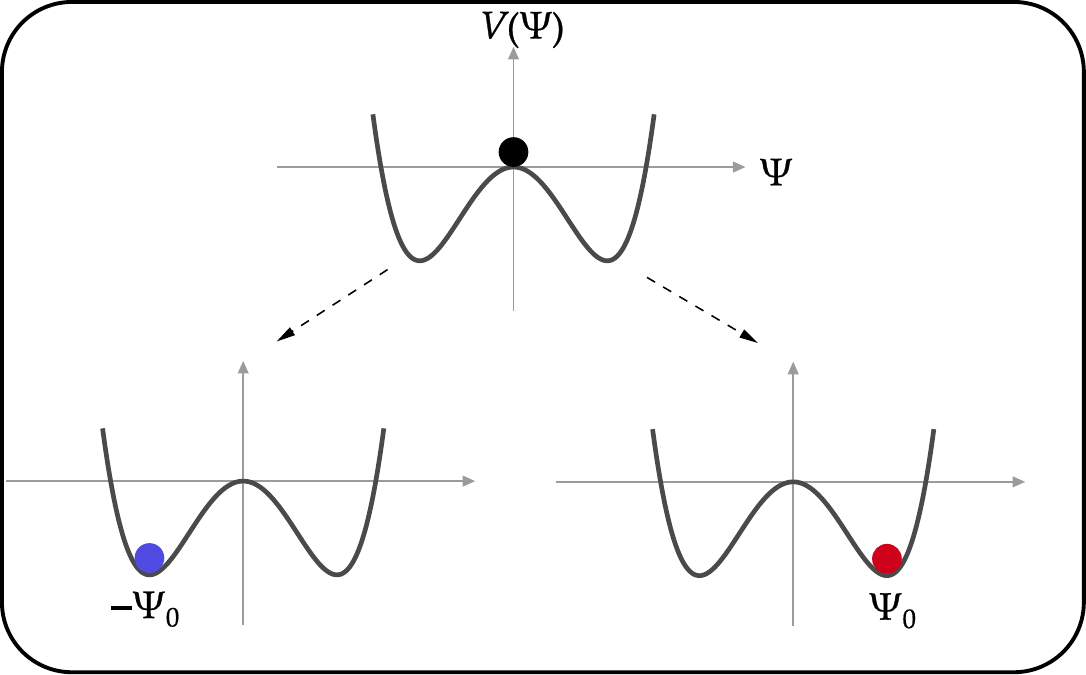}
\caption{The sketchy figure to illustrate the correspondence between the kink formations in the AdS bulk and in the AdS boundary field theory. The black hole indicates that this is a finite temperature phase transition. Different colors in AdS bulk implies domain wall-like structures which forms kinks normal to the radial direction. Interfaces between the balls with different arrows indicate the kinks in the AdS boundary. The inset figure illustrates the $Z_2$ symmetry breaking of the scalar fields.}\label{fig0}
\end{figure}

Here, we dynamically realize the kink hairs of real scalar fields from the parity symmetry ($Z_2$ symmetry) breaking in the AdS bulk, refer to Fig.\ref{fig0}. By decreasing the temperature of the system across the critical point, the former $Z_2$ symmetry of the scalar fields spontaneously break along a spatial direction. Then kinks form at the interfaces of the symmetry breaking domains. By fixing the final temperature lower than the critical temperature for a while, we see the stable structures of the kinks in the final equilibrium state. This indicates that these kink hairs are dynamically stable. 

From the AdS/CFT correspondence \cite{Maldacena:1997re,Gubser:1998bc,Witten:1998qj}, its holographic dual can mimic the kink formations in a strongly coupled one-dimensional chain. See Fig.\ref{fig0} for the illustration of this correspondence. The interfaces between the balls with different arrows in the AdS boundary indicate the structures of kinks (we do not distinguish between kinks and anti-kinks in this paper), which resembles the kink formation in a one-dimensional chain in condensed matter physics, such as the Ising chain \cite{suzuki}.  We find that the kink numbers in the boundary field theory satisfy a universal power-law relation to the quench rate, consistent with the KZM's prediction. In recent, it was proposed that the statistics of kink numbers satisfies a binomial distribution and the higher cumulants are predicted to obey a universal power-law scaling to the quench rate as well \cite{delCampo:2018hpn}. We examine this binomial distribution and power-law relation for the holographic kinks and find that the numerical results match the theoretical predictions very well. In addition, we investigate the probability of vanishing kinks, which can observe the onset of adiabaticity from the critical dynamics. The numerical fitting of this probability against the mean kink numbers matches the theoretical predictions closely.  Other holographic models of the topological defects from the continuous symmetry breaking, such as U(1) symmetry breaking can be found in \cite{Sonner:2014tca,Chesler:2014gya,Das:2014lda,Natsuume:2017jmu,Zeng:2019yhi,Xia:2020cjl,Li:2019oyz,delCampo:2021rak,Li:2021iph,Li:2021dwp,Xia:2021xap,Li:2021mtd,Li:2021jqk,delCampo:2022lqd,Zeng:2022hut}. The papers \cite{Das:2014lda,Natsuume:2017jmu} studied the KZ scalings in an analytical way and found consistent results with KZM, while other papers \cite{Sonner:2014tca,Chesler:2014gya,Zeng:2019yhi,Xia:2020cjl,Li:2019oyz,delCampo:2021rak,Li:2021iph,Li:2021dwp,Xia:2021xap,Li:2021mtd,Li:2021jqk,delCampo:2022lqd,Zeng:2022hut} numerically realized the formation of defects, either vortices in two-spatial dimensions or winding numbers in one-spatial dimension. However, they all dealt with complex scalar fields with U(1) symmetry, which is different from our paper in which we will study the real scalar fields with discrete $Z_2$ symmetry.


\section{Model of parity symmetry breaking}
\label{Z2}
 We start with a U(1) symmetric Abelian-Higgs model in the bulk with the Lagrangian \cite{Hartnoll:2008vx},
\be\label{action}
\mathcal{L}=-\frac14F_{\mu\nu}F^{\mu\nu}-|D_\mu\tilde\Psi|^2-m^2|\tilde\Psi|^2,
\ee
where $F_{\mu\nu}=\partial_\mu A_\nu-\partial_\nu A_\mu$ and $A_\mu$ is the U(1) gauge field.  $D_\mu=\nabla_\mu-iA_\mu$ and $\tilde\Psi$ is the complex scalar field. For simplicity, we work in the probe limit. Thus, the equations of motions (EoMs) read,
\be\label{eq1}
D_\mu D^\mu \tilde\Psi-m^2\tilde\Psi=0,\qquad  \nabla_\mu F^{\mu\nu}=J^\nu.
\ee
in which $J^\nu=i\left[\tilde\Psi^*D^\nu\tilde\Psi-\tilde\Psi(D^\nu\tilde\Psi)^*\right].$ In order to transform the Lagrangian with U(1) symmetry to $Z_2$ symmetry, we need to transform the complex scalar fields to real scalar fields. To this end, we make the following gauge transformations 
\be \tilde\Psi=\Psi e^{i\lambda},\qquad  A_\mu=M_\mu+\partial_\mu\lambda, \label{gaugefix}\ee
 where $\Psi, M_\mu$ and $\lambda$ are real functions. With these real functions, the previous Eq.\eqref{eq1} can be rewritten as 
\be\label{eq2}
\mathbb{D}_\mu\mathbb{D}^\mu\Psi-m^2\Psi=0, \qquad  \nabla_\mu F^{\mu\nu}=2M^\nu\Psi^2. 
\ee
where $\mathbb{D}_\mu=\nabla_\mu-iM_\mu$. (Alternatively, one can rewrite the action Eq.\eqref{action} in a St\"uckelberg form, then gauge the phase to be zero \cite{Franco:2009yz}. These two procedures are equivalent.) Therefore, if $+\Psi$ is a solution to the Eq.\eqref{eq2}, $-\Psi$ must be a solution as well. This indicates a $Z_2$ symmetry ($+\Psi\leftrightarrow -\Psi$) of the Eq.\eqref{eq2}. 



In an asymptotic AdS spacetime, as the temperature of the black hole $T$ is lower than a critical value $T_c$, the scalar fields will condensate near the black hole \cite{Gubser:2008px}. This idea is the prototype of the holographic superconductor/superfluid \cite{Hartnoll:2008vx}. Therefore, we see that $\pm\Psi$ are the $Z_2$ degenerated solutions to the ground state in the low temperature. That is, as $T<T_c$ the ground state will spontaneously break the $Z_2$ symmetry by choosing either $+\Psi$ or $-\Psi$. However, the above arguments goes only for static case. In a dynamical processing, quenching the system from high temperature ($T>T_c$) to low temperature ($T<T_c$), scalar fields will condensate in a different way. According to the KZM \cite{Kibble:1976sj,Kibble:1980mv,Zurek:1985qw}, dynamically driving the system to a symmetry breaking phase will induce the formation of topological defects in the spatial direction. Besides, these topological defects will be randomly distributed in space. In our case, kink formations will occur due to the parity symmetry breaking.

We adopt the idea of KZM to realize the kink formation in the bulk, i.e. the ``kink hairs" of a black hole in an asymptotic AdS spacetime, refer to Fig.\ref{fig0}.  In this case we need to solve the system in a dynamically. A convenient choice is to use the Eddington-Finkelstein coordinates in the planar Schwarzschild-AdS black hole \cite{Sonner:2014tca,Chesler:2014gya}, 
\be \label{metric}
ds^2=\frac{1}{z^2}\left[-f(z)dt^2-2dtdz+dx^2+dy^2\right].
\ee
in which $f(z)=1-(z/z_h)^3$ with $z_h$ the horizon position. Without loss of generality, we set the AdS radius be $\ell_{\rm AdS}=1$ and the horizon location be $z_h=1$ as well. Thus the AdS boundary is located at $z=0$ and the temperature of the black hole is $T=3/(4\pi)$. 

\section{Boundary conditions and numerical schemes}
\label{methods}
We set the ansatz of fields as 
$\Psi=\Psi(t,z,x),\ M_t=M_t(t,z,x), \ M_z=M_z(t,z,x),\ M_x=M_x(t,z,x)$,
and we have turned off the fields $M_y$. In this ansatz we assume a homogeneous dependence on the $y$-direction, besides we set the periodic boundary condition along $x$-direction.  Thus, the boundary field theory is effectively one-dimensional.  
The above ansatz of the fields is self-consistent with the EoMs \eqref{eq2} since there are four independent equations to solve four real functions. We should stress that the inclusion of $M_z$ is important since we are woking in the Eddington-Finkelstein coordinates. Even in the static case we cannot exclude $M_z$, instead $M_z=M_t/f(z)$ in static. See the Appendix \ref{appa} for details. 

The expansions of the fields near AdS boundary $z=0$ is (for simplicity we set $m^2=-2$), 
$\Psi\sim\Psi_1(t,x)z+\Psi_2(t,x)z^2+\mathcal{O}(z^3), \quad M_t\sim \mu(t,x)-\rho(t,x)z+\mathcal{O}(z^3),
 M_z\sim a_z(t,x)+b_z(t,x)z+\mathcal{O}(z^3), \quad M_x\sim a_x(t,x)+b_x(t,x)z+\mathcal{O}(z^3)$.  
We choose the standard quantization by setting $\Psi_1(t,x)\equiv0$, thus $\Psi_2$ is related to the condensate of the superconducting order parameter $O(t,x)$ in the boundary field theory. $\mu$ and $\rho$ are interpreted as the chemical potential and charge density respectively in the boundary.  From the expansions of the equation \eqref{eq2} we find that $a_z=\mu$.  $a_x$ and $b_x$ are the velocity and current of the gauge fields on the boundary. We set $a_x=0$ to work in the holographic superfluid model. Near the horizon we set $M_t=0$ as usual, and let other fields be finite.

In the KZM, one needs to quench the system across the critical point which is obtained from the equilibrium states. In this paper we quench the chemical potential $\mu$. In the equilibrium or in the static case we find the critical potential is $\mu_c\approx 4.06$. We linearly quench the system from $T=1.4T_c$ to $T=0.8T_c$ and the linear quench profile is 
\be T(t)/T_c=1-t/\tau_Q, \label{quench}\ee
 where $\tau_Q$ is the quench rate. From the dimensional analysis we know that $[T]=[\mu]=[{\rm mass}]$, thus $T/\mu$ is massless. From the setup of holographic superconductor \cite{Hartnoll:2008vx} we see that decreasing $T$ is equivalent to increasing $\mu$, i.e., $T(t)/T_c=\mu_c/\mu(t)$. Therefore, we quench the chemical potential as $\mu(t)=\mu_c/(1-t/\tau_Q)$. 

At initial time, the system is in a state with vanishing scalar fields. In this case the simple solution to the gauge fields are $M_t=M_zf(z)=\mu(t_i)-\mu(t_i)z$ and $M_x=0$. However, in order to quench the system and evolve it to a state with topological defects (kinks in our paper), a common procedure is to introduce white noise into the system at initial time. To this end, we introduce a very small Gaussian white noise $\zeta(x_i,t)$ for the scalar fields in the bulk with $\langle\zeta(x_i,t)\rangle=0$ and $\langle\zeta(x_i,t)\zeta(x_j,t')\rangle=h\delta(t-t')\delta(x_i-x_j)$, in which $h=0.001$. Then we thermalize this system by fixing the temperature at $T=1.4T_c$ for a while in order to diminish the extra virtual kinks caused by the noise. After the thermalization we quench the system according to Eq.\eqref{quench}. The system will evolve according to the Eqs.\eqref{eq2}. Once the temperature arrives at $T=0.8T_c$ we stop the quench and maintain the temperature until the system reach the final equilibrium.  We adopt the 4th Runge-Kutta method in the time direction with time step $\Delta t=0.01$. In the radial direction $z$, we use the Chebyshev pseudospectral methods with $21$ grid points. In the periodic $x$-direction, we use the Fourier decomposition with 401 grid points. The length of the loop along $x$-direction is set to be $L=200$.


\begin{figure}[t]
\centering
\includegraphics[trim=3.5cm 9.cm 2cm 10cm, clip=true, scale=0.47]{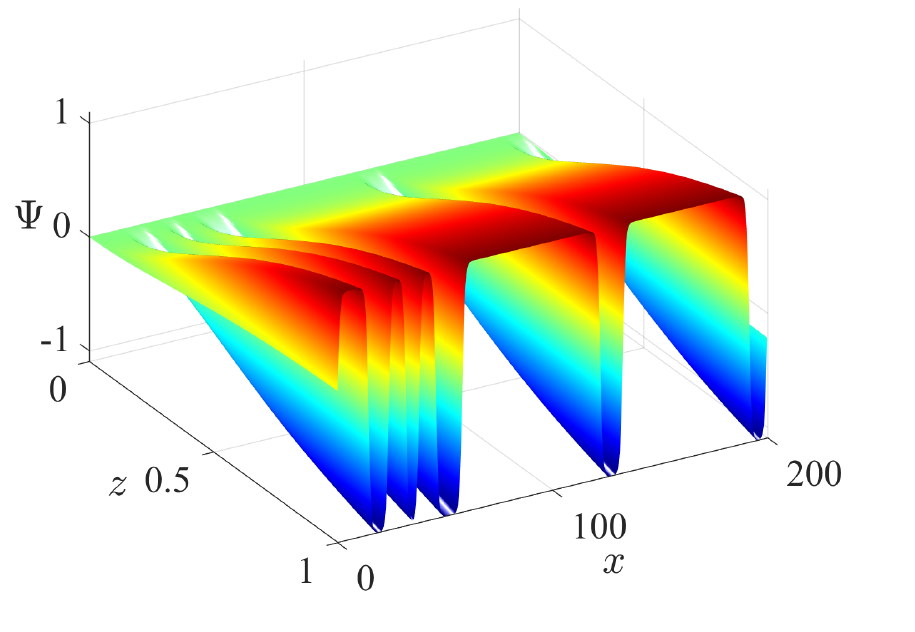}
\put(-200,130){$\bf a$}
\includegraphics[trim=3.25cm 9.cm 4cm 9.9cm, clip=true, scale=0.45]{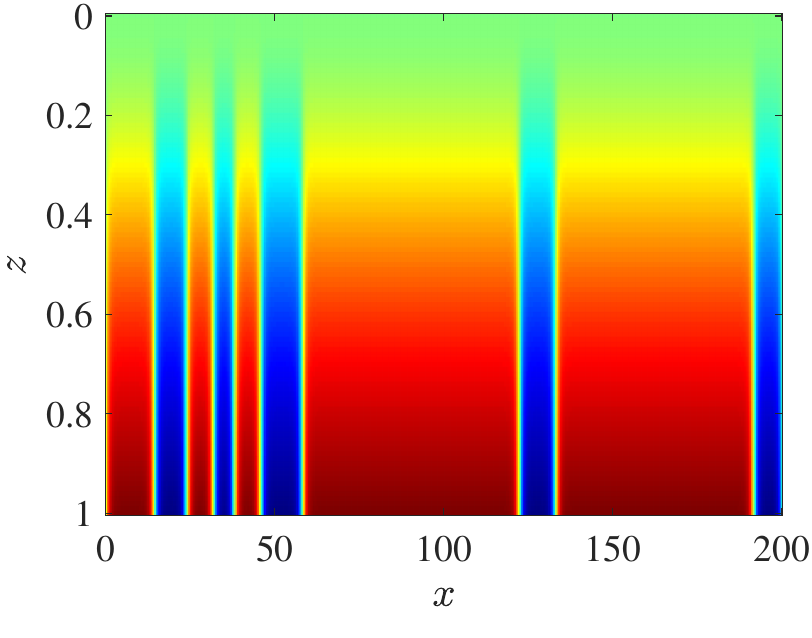}
\put(-180,130){$\bf b$}
\\
\quad\includegraphics[trim=3.cm 9.cm 3.5cm 9.5cm, clip=true, scale=0.46]{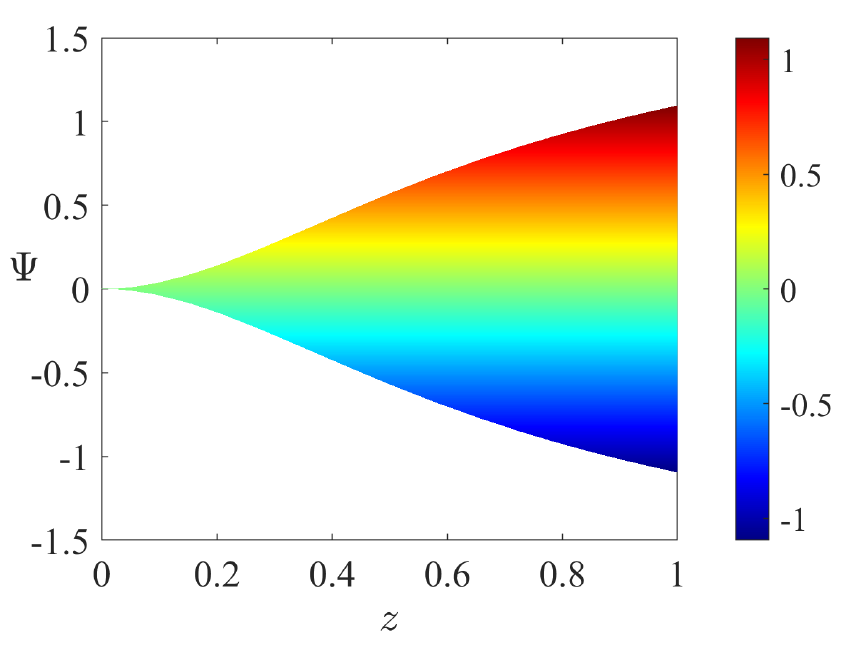}
\put(-190,130){$\bf c$}
\includegraphics[trim=2.cm 9.cm 3cm 9.5cm, clip=true, scale=0.46]{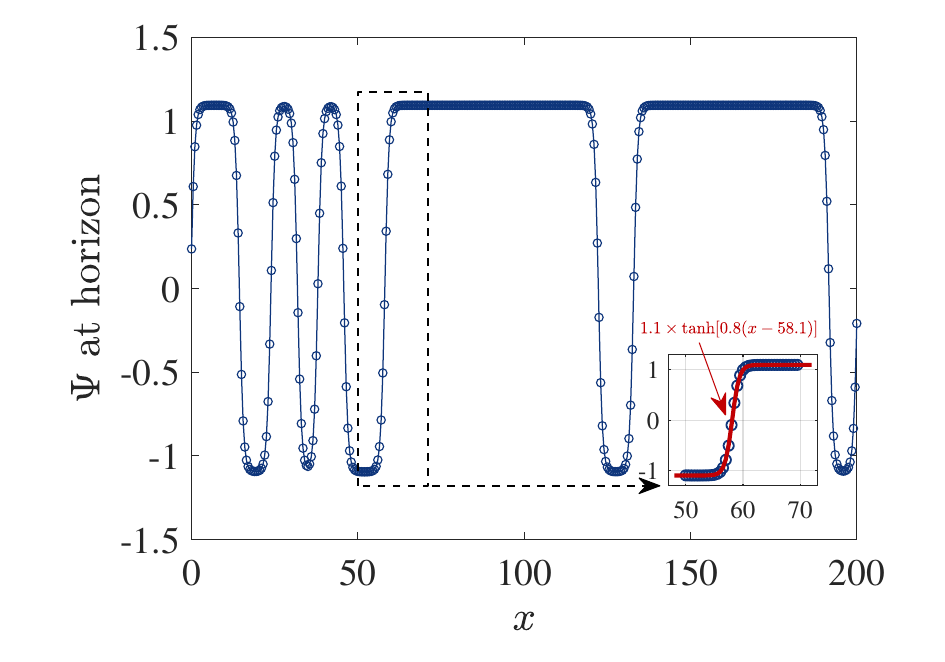}
\put(-190,130){$\bf d$}
\caption{Kink formations in the AdS bulk. ({\bf a}) Stereoscopic visualization of the kink formations in the AdS bulk; ({\bf b}) Density plot of the kink formation in the bulk; ({\bf c}) Profile of the kink formation in the bulk from the viewpoint parallel to the $x$-direction; ({\bf d}) Profile of the kinks at the horizon. The inset plot shows the fitted line (in red) of a kink in the dashed box.  }\label{fig1}
\end{figure}

\section{Results}
\subsection{Kink hairs in the AdS bulk}
 Quench the system through the critical point $T_c$ from high temperature to low temperature (refer to Eq.\eqref{quench}), the system will enter a superconducting phase and the formation of kinks will turn out due to the KZM \cite{Kibble:1976sj,Kibble:1980mv,Zurek:1985qw}. 
In Fig.\ref{fig1} we exhibit the kink formations in the AdS bulk at the final equilibrium state.  From the panel (a) of Fig.\ref{fig1}, we see that the kink structures exist along the $x$-direction while in the $z$-direction they perform mildly from the horizon $(z=1)$ to the boundary $(z=0)$.\footnote{In order not to make any confusion, we should stress that the structures in the bulk look like domain walls. However, it is safe to call them `kinks' since the kink structures only exist along $x$-direction and no kink structures exist along $z$-direction. Therefore, the topological structure of kinks `effectively' exist in one-dimensional space.}  Panel (b) shows the densities of the scalar fields in the bulk, from which we can see the kink structures clearly along the $x$-direction. It is found that some kinks are broader and others are narrower. This is due to the random distributions of the kinks.  
Panel (c) exhibits the profile of the kinks from the view parallel to $x$-direction. The profile is perfectly symmetric along $\Psi=0$ axis, indicates the $Z_2$ symmetry breaking in the bulk along the radial direction. 
From panel (d) we can see the one-dimensional kinks at the horizon. The absolute value of the maximum are equal to the absolute values of the  minimum of the kinks, which is the reflection of the $Z_2$ symmetry breaking. In the mean field theory one single kink behaves as $\Psi=\Psi_{\rm max} \tanh[a(x-b)]$, where $a$ is proportional to the mass of the scalar field while $b$ is the locations of the kink \cite{Weinberg:1992hc}. In the inset plot of panel (d), we show the the profile of a single kink in the dashed box at round $x\approx 58.1$ and its best fit (red line) as $1.1\times\tanh[0.8(x-58.1)]$. We see that the theoretical prediction and the numerical results match very well. 
Therefore we realize the ``kink hairs" from $Z_2$ symmetry breaking near the horizon and also in the bulk.
 




\subsection{Holographic kink formation in the AdS boundary}
From the holographic superconductor \cite{Hartnoll:2008vx}, the scalar hairs in the bulk will induce the condensate of the order parameters in the boundary field theory. Thus, we speculate that the kinks of scalar fields in the bulk will also induce the kinks of the order parameter in the boundary field theory which is strongly coupled. 

\begin{figure}[t]
\centering
\includegraphics[trim=0.5cm 0.cm 5.8cm 1.3cm, clip=true, scale=0.45]{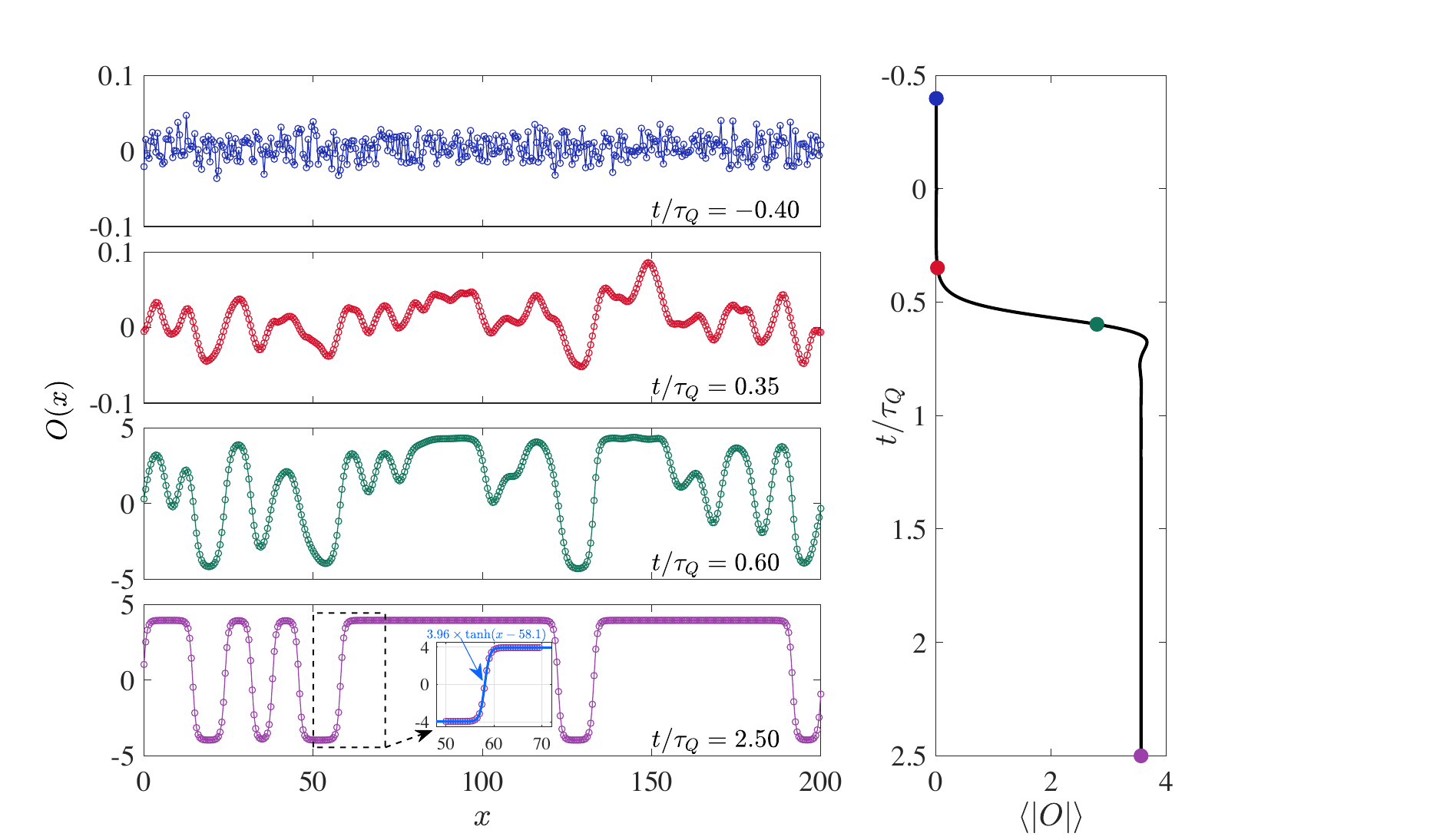}
\put(-315,205){$\bf a$}
\put(-60,205){$\bf b$}
\caption{Kink formations in the AdS boundary. ({\bf a}) Snapshots of the order parameter $O(x)$ at the AdS boundary along the $x$-direction at four different times $t/\tau_Q=-0.40$, $0.35$, $0.60$ and $2.50$ with the quench rate $\tau_Q=20$. The inset plot in the lowest panel exhibits the best fit (in blue) of a kink in the dashed box;   ({\bf b}) Time evolution of the average absolute values of the order parameter $\langle|O|\rangle$. Four colored points correspond to the four time stages in the panel ({\bf a}). }\label{fig2}
\end{figure}


In Fig.\ref{fig2} we exhibit the kink formations of the order parameter $O(x)$ in the boundary field theory corresponding to the kink hairs in the Fig.\ref{fig1}.  In the panel (a), we show the snapshots of the order parameter along the $x$-direction at four different time stages, i.e., from the initial to the final equilibrium state. The quench rate in this figure is $\tau_Q=20$. At the initial time $t/\tau_Q=-0.4$, the order parameter exhibits a random profile and then evolves and grows rapidly around $t/\tau_Q=0.35$. At the time $t/\tau_Q=0.6$ the order parameter are well developed but still not in the equilibrium state. We stop the evolution at $t/\tau_Q=2.5$, when the system arrives at the final equilibrium state and the order parameter develops stable kinks. Compare the time $t/\tau_Q=0.6$ and $t/\tau_Q=2.5$ we see that some kinks do not change their locations (for instance the kink at around $x=140$) and some minor ripples disappear due to the coarsening dynamics. The inset plot in the last snapshots shows the kink at around $x=58.1$, and its best fit in blue $3.96\times\tanh(x-58.1)$, which matches the theoretical prediction very well. Compare to the panel (d) in Fig.\ref{fig1} we see that except the differences of the amplitudes and the slopes, the locations of kinks at the horizon and at the $z=0$ boundary do not change. 
Panel (b) of Fig.\ref{fig2} shows the time evolution of the corresponding average absolute values of the order parameter $\langle|O|\rangle$ from the initial time to the final equilibrium time where it arrives at a plateau.  


\subsection{Kink number distribution and universal scalings}

We need to note that the kink numbers always appear as even because of the periodic boundary conditions we adopt \cite{delCampo:2021rak}. Therefore, we count the number of kink pairs $n$ in the following rather than counting the number of individual kinks.  It is useful to examine whether the kink formations in the holographic model satisfies the KZM's scaling relation between the average numbers $\langle n\rangle$ and the quench rate $\tau_Q$,
\be\label{kzscaling}
\langle n\rangle \propto {\tau_Q}^{-{d\nu}/{(1+z\nu)}},
\ee
where $d$ is the effective spatial dimension of the topological defects, $\nu$ and $z$ are the static critical exponent and dynamic critical exponent in the equilibrium, respectively \cite{Zurek:1985qw}. It is noted that $d=1$ in our case both for the kinks in the bulk and those in the AdS boundary, since the kinks form only along the $x$-direction while other spatial directions do not have the structures of defects. Besides, from the analysis in Fig.\ref{fig1} and Fig.\ref{fig2} we see that the kink numbers in the bulk are equivalent to those in the boundary. As reported previously \cite{Sonner:2014tca,Maeda:2009wv,Zeng:2019yhi}, the boundary field theory is a mean field theory therefore we have $\nu=1/2$ and $z=2$.  


\begin{figure}[t]
\centering
\includegraphics[trim=1.5cm 0.3cm 1.2cm 1.8cm, clip=true, scale=0.5]{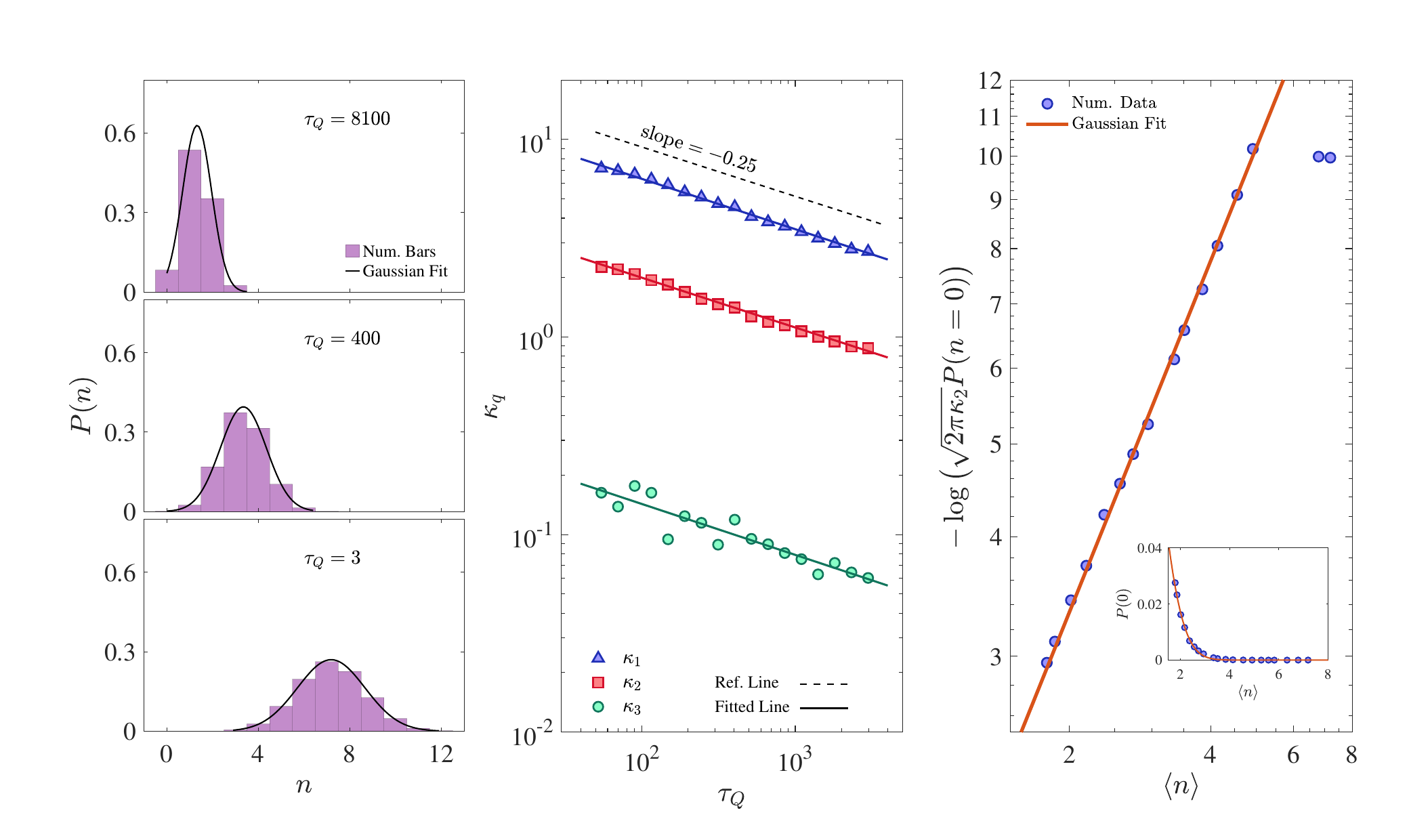}
\put(-445,265){\bf a}
\put(-295,265){\bf b}
\put(-145,265){\bf c}
\caption{({\bf a}) Histogram of the probability density of kink pair numbers with different quench rates. The solid lines are the best fit with Gaussian distributions; ({\bf b}) Double logarithmic plot of the first three cumulants $\kappa_q$ to the quench rate $\tau_Q$. In the slow quench regime (greater $\tau_Q$), the cumulants satisfy a power law with a common power indicated by the almost parallel fitted lines; ({\bf c}) Double logarithmic plot between $-\log\left(\sqrt{2\pi\kappa_2}P(n=0)\right)$ and the mean kink numbers $\langle n\rangle$. The fitted line indicates that they satisfy the Gaussian approximations Eq.\eqref{gau} very well. The inset plot is the relation between $P(n=0)$ and $\langle n\rangle$ in the normal axis.  }\label{fig3}
\end{figure}

Beyond the KZM, distributions of kink numbers have been theoretically investigated beyond the mean density. The kink numbers are assumed to satisfy the binomial distribution which can be associated with independent Bernoulli trials \cite{delCampo:2018hpn}. The first three cumulants of the kink pair numbers, i.e., $\kappa_1=\langle n\rangle, \kappa_2=\langle(n-\langle n\rangle)^2\rangle $ and $\kappa_3=\langle(n-\langle n\rangle)^3\rangle$ are expected obey a universal power-law scaling as $\kappa_{q}\propto\tau_Q^{-{d\nu}/{(1+z\nu)}}, (q=1,2,3)$. The ratios between the cumulants can be readily obtained from the binomial distributions as 
\be \label{k2k1}\kappa_2/\kappa_1&=&(2-\sqrt2)/2\approx 0.293, \\
\label{k3k1} \kappa_3/\kappa_1&=&1-3/\sqrt2+2/\sqrt3\approx 0.033. \ee
Moreover, from the central limit theorem the binomial distribution can be approximated by the Gaussian distribution in the large limit of trial numbers with fixed probability of success. Therefore, the probability density of the kink number $P(n)$ can be approximated as
\be\label{gau} P(n)\approx \frac{1}{\sqrt{2\pi\kappa_2}}\exp\left[-\frac{(n-\langle n\rangle)^2}{2\kappa_2}\right]\ee

In Fig.\ref{fig3}(a), the histogram of the probability density of the kink pair numbers $P(n)$ are exhibited. For the three different quench rates, i.e, $\tau_Q=8100, 400$ and $3$, the probability density can be well approximated by the Gaussian distribution (black solid lines) in Eq.\eqref{gau}.   In Fig.\ref{fig3}(b) we show the relations between the first three cumulants $\kappa_{1,2,3}$ to the quench rate $\tau_Q$. We have simulated independently $10000$ times to generate this figure. 
In the slow quench regime as $\tau_Q$ is greater, the three cumulants satisfy a universal power law to the quench rate with a roughly the same power as the fitted lines indicate. Specifically, the fitted lines are $\kappa_1\approx (20.48\pm0.12)\times\tau_Q^{-0.256\pm0.012}$, $\kappa_2\approx (6.38\pm0.06)\times\tau_Q^{-0.256\pm0.015}$ and $\kappa_3\approx (0.48\pm0.09)\times\tau_Q^{-0.258\pm0.054}$.   Therefore, the numerical power is consistent with the theoretical prediction $-d\nu/(1+z\nu)=-0.25$ in Eq.\eqref{kzscaling}. From the fittings, the numerical ratios of these cumulants are $\kappa_2/\kappa_1\approx 0.312$ and $\kappa_3/\kappa_1\approx0.023$. Compared to the theoretical predictions in Eqs.\eqref{k2k1} and \eqref{k3k1} we see that the ratio $\kappa_2/\kappa_1$ is close to the theoretical predictions, however, the ratio $\kappa_3/\kappa_1$ has roughly $30\%$ deviations from the theoretical prediction. This deviation is because the numerical statistics of $\kappa_3$ relies much on the amount of the simulation data \cite{Guo,delCampo:2021rak}. More simulations of $\kappa_3$ are expected to improve this situation. 

Probability of vanishing kinks $P(n=0)$ reveals the onset of adiabaticity of the critical dynamics. From the Eq.\eqref{gau} we can estimate the relations between $P(n=0)$ against the mean numbers $\langle n\rangle$. In the Fig.\eqref{fig3}(c) we show the double logarithmic plot between $P(n=0)$ and $\langle n\rangle$, and observe that they indeed satisfy the Gaussian approximations Eq.\eqref{gau}.





\section{Discussions}
We have initiated the study of kink hairs in the AdS black hole due to the spontaneous $Z_2$ symmetry breaking of the real scalar fields in the bulk. By quenching the temperature of the system across the critical point, the kink hairs turned out in the bulk as a result of the KZM. This is a counter-example to the {\it no hair conjecture} in black hole.   Holographically the kinks in the bulk induced the kink formation in the boundary field theory, which could model the kinks in a one-dimensional chain in condensed matter physics. Since the kinks appeared in the form of kinks and anti-kinks, we made the full counting statistics of the pair numbers of kinks. The mean numbers of the kink pairs were found to satisfy a power-law scaling to the quench rate, consistent with the KZM. Moreover, beyond the KZM, the number distributions of kinks were assumed to be binomial distribution and higher cumulants were found to be proportional to the mean numbers.  The ratios between these cumulants were found to be consistent with theoretical predictions, although $\kappa_3/\kappa_1$ is a little bit away from theories. We gave some explanations of this deviation since $\kappa_3$ depended much on the simulation times. This deviation was also observed in some existing literatures \cite{Guo,delCampo:2021rak}. In the large limit of the trial numbers, the binomial distribution can be approximated by the Gaussian distributions. We indeed found that the histogram of the probability density of the kink pair numbers could be well fitted by the Gaussian distribution. Furthermore, this Gaussian distribution could also predict the probability of vanishing kinks, which could be used to study the adiabaticity limit of the critical dynamics. We found that the numerical results of the probability of vanishing kinks matched the theoretical predictions very well. 

 From the AdS/CFT correspondence the boundary field theory is a mean field theory \cite{Zaanen:2015oix}, thus, our model compares to the Ginzburg-Landau theory in studying the kinks in one-dimensional chain \cite{Laguna:1997cf,Gomez-Ruiz:2019hdw}.  In the Appendix \ref{appb} of this paper, we have shown the kink formations due to the spontaneous $Z_2$ symmetry breaking of real field $\phi$ in the time-dependent Ginzburg-Landau (TDGL) theory. Compared to this TDGL theory, we found that the holographic model had similar behaviors to the TDGL model. In particular, the universal scalings between the cumulants of the kink pair numbers and the quench rates all had the scaling powers close to $-0.25$, consistent with the KZM predictions in mean field theory. Furthermore, for the TDGL model the ratios between the cumulants were also very close to the theoretical predictions, demonstrating that the formations of kinks satisfy the binomial distributions \cite{delCampo:2018hpn}. Interestingly, the original assumption of the binomial distribution of kink formations as well as the universal scalings between the cumulants and the quench rate were made in a quantum Ising model with transverse magnetic fields, however, here we found that this assumption could also be applied in the holographic model and TDGL model. From this sense, we can argue that our holographic kink formations is a kind of ``effective model" to mimic the Ising model or TDGL model in condensed matter physics. In the Appendix \ref{appb} we also studied the Gaussian approximations of the probability distribution for the kinks just like we did in the holographic studies, and we found that for the TDGL model the numerical results matched the Gaussian approximations very well. In the adiabatic limit, i.e., in the limit of vanishing kinks, the probability distribution could also be approximated by the Gaussian distribution in the TDGL model. For details of the studying of TDGL model, please refer to the Appendix \ref{appb}.  However, the merit of our holographic model is that the boundary field theory is strongly coupled while the TDGL theory is weakly coupled \cite{Zaanen:2015oix}. Findings in this paper indicate that the strong coupling does not change the universal power-law scalings between the first three cumulants to the quench rate compared to weak theory. And the probabilities of the vanishing kinks also match the Gaussian approximations in the weak theory. 

A caveat is that the one-dimensional Ising chain does not possess any finite temperature phase transition, however it has phase transition from paramagnetic to ferromagnetic if one drives the external transverse magnetic field across the critical point \cite{suzuki}. Therefore, our holographic model of the kinks formation resembles the one-dimensional Ising chain in such a sense that the temperature in our case is similar to the magnetic field strength in Ising chain. However, the striking difference between the mean field theory (including holographic model and TDGL model) and the quantum Ising model resides in the distinct powers of the universal scalings between the cumulants and the quench rate.  In the mean field theory this power is $-0.25$, while for the quantum Ising model with transverse magnetic field it is $-0.5$ \cite{delCampo:2018hpn}. This difference was due to the different dynamical critical exponents $z$ and static critical exponents $\nu$ in these two models. For mean field theory they are $z=2$ and $\nu=1/2$, thus from the Eq.\eqref{kzscaling} the power is $-0.25$; while for Ising model they are $z=\nu=1$ thus the power is $-0.5$. Although the scaling powers are distinct between mean field theory and the Ising model, the ratios between the cumulants are similar and match the theoretical predictions from the Ising model (refer to Eqs.\eqref{k2k1} and \eqref{k3k1}). As we have already discussed above, this indicates that the kink formations in the holographic model and the TDGL model also satisfy the binomial distributions, which was previously proposed to be the distributions of kinks in the Ising model. In this sense we can say that the mean field theory model is an ``effective model" for the Ising model.  

For future prospects, there also exists structural phase transitions from linear to a doubly-degenerate zigzag phases, such as the trapped ion chains \cite{Retzker,delcampo2010}. Our holographic model may have relevances to such kind of phase transition as well. 
This holographic model of kinks in one-dimensional space can be readily extended to two-dimensional space and many interesting properties can be further studied, such as kink spatial distributions \cite{delCampo:2022lqd},  the entanglement entropy \cite{its} and etc. We expect that our initial work on the holographic kinks will shed light on studies in the strongly coupled topological defects.

\section*{Acknowledgements}

This work was partially supported by the National Natural Science Foundation of China (Grants No.11875095 and 12175008).

\appendix
\setcounter{equation}{0}
\setcounter{figure}{0}
\setcounter{table}{0}
\setcounter{section}{0}
\renewcommand{\theequation}{S\arabic{equation}}
\renewcommand{\thefigure}{S\arabic{figure}}

\section{Explicit forms of equations of motions}
\label{appa}
The EoMs \eqref{eq2} in the main text can be decomposed into 
\be\label{eq_app}
\nabla_\mu\nabla^\mu\Psi-M_\mu M^\mu\Psi-m^2\Psi&=&0, \label{eqapp1}\\
\left(\nabla_\mu M^\mu\right)\Psi+2M^\mu\nabla_\mu \Psi&=&0,\label{eqapp2}\\
\nabla_\mu F^{\mu\nu}&=&2M^\nu\Psi^2. \label{eqapp3}
\ee
The Eqs.\eqref{eqapp1} and \eqref{eqapp2} are respectively from the real part and imaginary part of the scalar equations in \eqref{eq2}. As was emphasized in \cite{Hu:2015dnl}, these equations are not independent. In fact, from the EoMs of gauge fields \eqref{eqapp3} one can derive the imaginary part EoMs \eqref{eqapp2}, such as
\be \label{cons}
0\equiv\nabla_\nu\left(\nabla_\mu F^{\mu\nu}\right)\Rightarrow \nabla_\nu\left(2M^\nu\Psi^2\right)=0  \Rightarrow  \left(\nabla_\nu M^\nu\right)\Psi+2M^\nu\nabla_\nu \Psi=0. 
\ee
The last equality is exactly the imaginary part of the scalar EoMs \eqref{eqapp2}.

In our ansatz of the fields and in the frame of the line-element Eq.\eqref{metric} in the main text, the above EoMs become \\
1). The gauge fields \eqref{eqapp3} part:
\be
0&=&-\frac{2\Psi^2M_t}{z^2}+\partial^2_xM_t+f\partial^2_zM_t-\partial_{tx}M_x-\partial_{tz}M_t-f\partial_{tz}M_z+\partial^2_tM_z,\label{em1}\\
0&=&-\frac{2\Psi^2M_z}{z^2}+\partial^2_xM_z-\partial_{zx}M_x+\partial^2_zM_t-\partial_{tz}M_z,\label{em2}\\
0&=&-\frac{2\Psi^2M_x}{z^2}-f'\partial_xM_z+f'\partial_zM_x+\partial_{zx}M_t-f\partial_{zx}M_z+f\partial^2_zM_x+\partial_{tx}M_z-2\partial_{tz}M_x,\label{em3}
\ee
2). The real part of scalar fields \eqref{eqapp1}: 
\be
0&=&-\frac{m^2\Psi}{2z^2}-\frac12\Psi M_x^2+\Psi M_t M_z-\frac12\Psi f M_z^2+\frac12\partial^2_x\Psi-\frac{f\partial_z\Psi}{z}+\frac12\partial_z(f\partial_z\Psi)
+\frac{\partial_t\Psi}{z}-\partial_{tz}\Psi
\ee
3). The imaginary part of scalar fields \eqref{eqapp2}:
\be
0&=&\frac{2\Psi}{z}(fM_z-M_t)-\Psi\left(\partial_z(fM_z)+\partial_x M_x-\partial_tM_z-\partial_zM_t\right) \nonumber\\
&&+2\left(M_t\partial_z\Psi+M_z\partial_t\Psi-M_x\partial_x\Psi-fM_z\partial_z\Psi\right)
\ee
in which $f'=f'(z)$. There are five equations, but only four of them are independent due to the constraint \eqref{cons}. Therefore, there are four independent equations for four real fields, i.e., $\Psi, M_t, M_z$ and $M_x$. Thus, our ansatz of the fields are self-consistent.  
We need to stress that including the $(t, x)$-dependence, our ansatz is the only possible choice. One cannot omit $M_z$ or $M_x$. 

In order to get the initial condition for our quench, we need to solve the static as well as $x$-independent case of the EoMs. Therefore, the EoMs \eqref{em1},\eqref{em2} and \eqref{em3} become
\be 
0&=&-\frac{2\Psi^2M_t}{z^2}+f\partial^2_zM_t,\label{em11}\\
0&=&-\frac{2\Psi^2M_z}{z^2}+\partial^2_zM_t,\label{em22}\\
0&=&-\frac{2\Psi^2M_x}{z^2}+f'\partial_zM_x+f\partial^2_zM_x. \label{em33}
\ee
From Eq.\eqref{em33} we can safely set $M_x=0$. From \eqref{em11} and \eqref{em22} we can readily get $M_z=M_t/f$. At the initial time the system is in the normal state with vanishing scalar fields $\Psi=0$. Thus we can solve $M_t=\mu-\mu z$ and $M_z=(\mu-\mu z)/f$ by imposing the boundary condition $M_t(z\to0)=\mu$ and $M_t(z\to1)=0$. 

\section{Kink formations in a time-dependent Ginzburg-Landau model}
\label{appb}
We start with a model of real non-conserved scalar order parameters $\phi$, which can be described by the time-dependent Ginzburg-Landau (TDGL) equation coupled with a Langevin noise term $\theta(t,\vec x)$ \cite{Laguna:1997cf}, 
\be\label{GLeq}
\ddot\phi+\eta\dot\phi-c^2\nabla^2\phi+\frac12\left(\beta\phi^3-m^2\epsilon(t)\phi\right)=\theta(t,\vec x),
\ee
where $\eta$ represents the viscosity, $c,\beta, m$ are constant coefficients and $\epsilon(t)$ is a relative distance to the critical point. The noise satisfies $\langle \theta(t,\vec x)\rangle=0$ and $\langle \theta(t',\vec x')\theta(t,\vec x)\rangle=2\eta T\delta(\vec x'-\vec x)\delta(t'-t)$ in which $T$ represents the temperature of the reservoir which will be kept constant in our paper. After the transformation, 
\be
t\to\frac tm,~\vec x\to\vec x\frac cm,~\eta\to\eta m,~\phi\to\phi\frac{m}{\sqrt\beta},~T\to T\frac{m^3c}{\beta},
\ee
the above Eq.\eqref{GLeq} can be transformed to a familiar form as
\be\label{GLeq2}
\ddot\phi+\eta\dot\phi-\nabla^2\phi+\frac12\left(\phi^3-\epsilon(t)\phi\right)=\theta(t,\vec x). 
\ee
It is well-known that in the TDGL model, as $\epsilon<0$ the solution $\phi=0$ is preferred, however as $\epsilon>0$ the preferred solution is $\phi\neq0$. Therefore, if we quench $\epsilon$ from negative, across the critical point $\epsilon=0$, to positive, the previous $Z_2$ symmetries of the real scalar field $\phi$ will spontaneously break and the kinks will form. For simplicity, we linearly quench $\epsilon$ as $\epsilon(t)={t}/{\tau_Q}$, 
where $\tau_Q$ is the quench rate. Therefore, we can quench the system from negative $\epsilon$ to positive $\epsilon$. Specifically, we start the quench from $t/\tau_Q=-1.6$ to $t/\tau_Q=10$, then we stop the quench and keep it at $\epsilon_f=10$. Until the final equilibrium state, the order parameter will take the values $\phi=\pm\sqrt{\epsilon_f}=\pm\sqrt{10}$.  To analytically solve the Eq.\eqref{GLeq2} is formidable, thus we resort to numerical solutions. In the numerics, we have set $\eta=1$ and $T=0.01$. The length of the system is $L=200$ with the periodic boundary conditions. Therefore, in the spatial direction we Fourier decompose the length to $10^3$ grids. In the time directions, the 4th Runge-Kutta methods are adopted with the time step $\Delta t=0.01$. 

\begin{figure}[h]
\centering
\includegraphics[trim=1.cm 0cm 5.5cm 1cm, clip=true, scale=0.5]{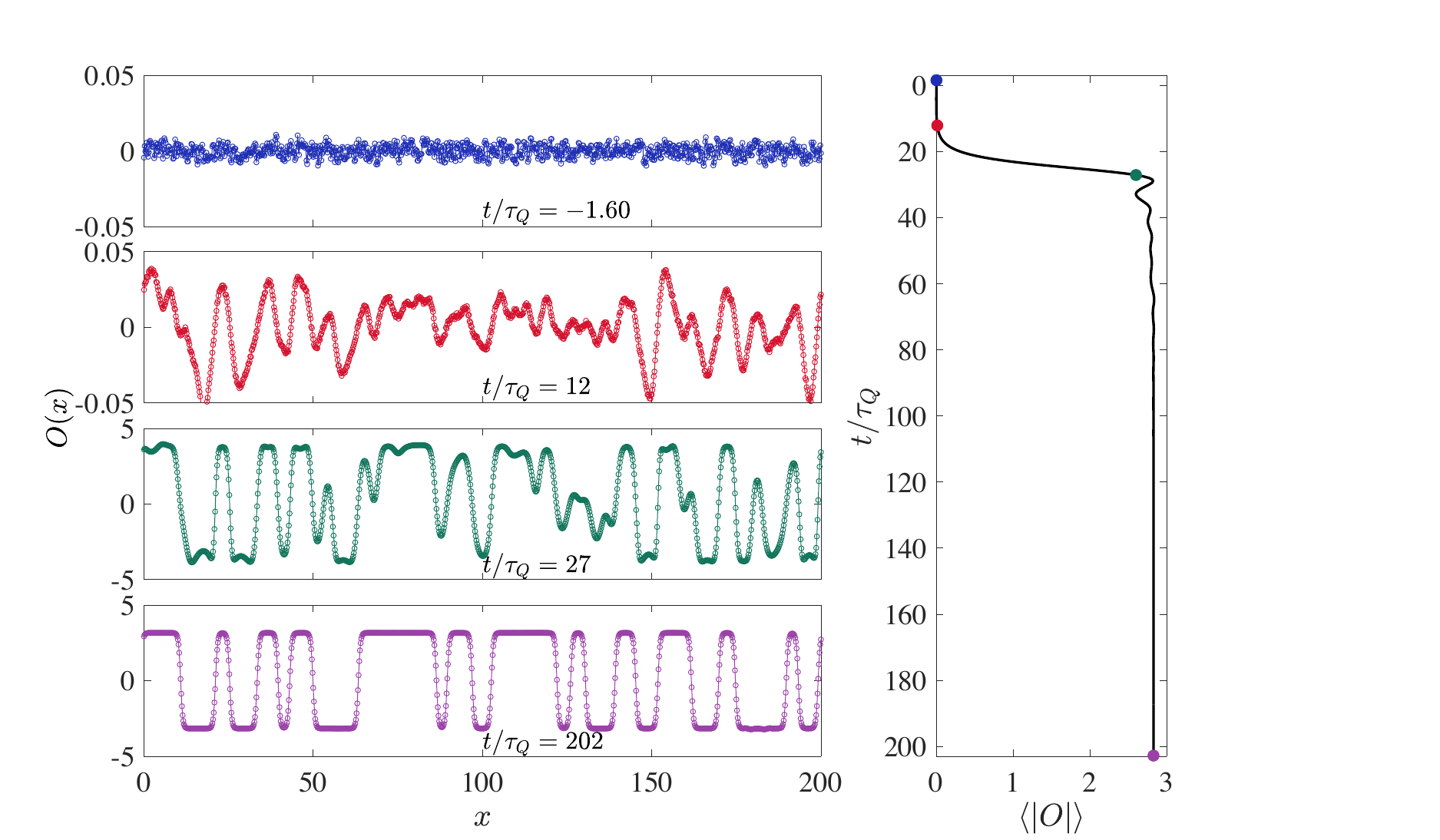}
\put(-355,235){\bf a}
\put(-100,235){\bf b}
\caption{Kink formations in the TDGL model. ({\bf a}) Snap shots of the kink configurations at four different stages with the quench rate $\tau_Q=e^1$; ({\bf b}) Time evolution of the average absolute values of the order parameter. The colored points correspond to the four snap shots in panel ({\bf a}). }\label{GL_evol}
\end{figure}

From Fig.\ref{GL_evol}(a) we see the four snap shots of the time evolution of the kink profile from the initial time to the final equilibrium state. At $t/\tau_Q=-1.6$ the scalar field is random with very tiny amplitudes which will serves as inhomogeneous seeds for the time evolution of the system. At $t/\tau_Q=12$ the scalar field is still random but with larger amplitudes. It is the instant before the rapid growth of the average order parameter, see the red point in panel (b). Later at $t/\tau_Q=27$ the system already underwent the rapid growth and just before the plateau of the average order parameter, see the green point in panel (b). At this instant, the amplitudes of the scalar fields already developed, but the kinks are not well formed. At the final equilibrium time $t/\tau_Q=202$ the kinks are well developed and they will stay in this form for a long time, which are dynamically stable. 

\begin{figure}[h]
\centering
\includegraphics[trim=3.4cm 9.3cm 3.4cm 9.5cm, clip=true, scale=0.8]{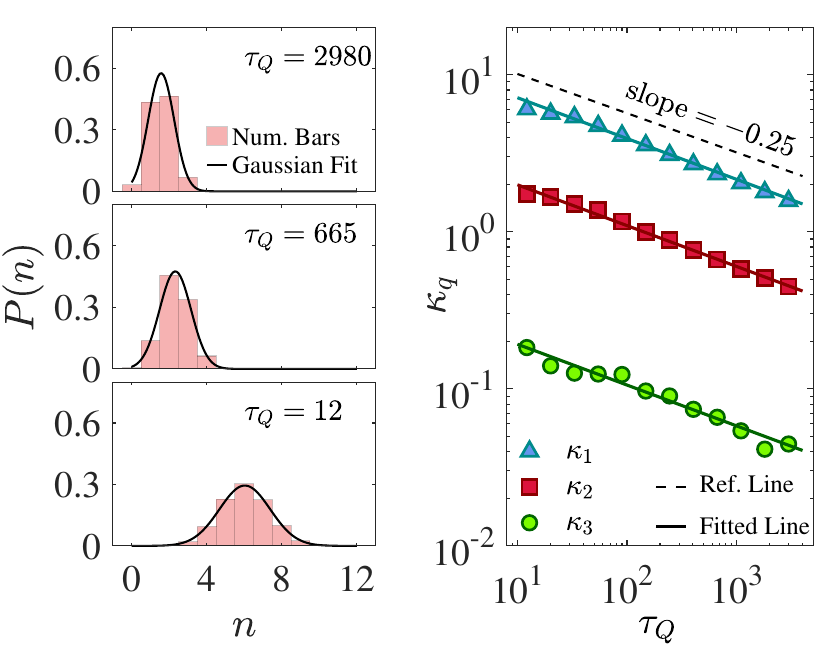}
\put(-295,235){\bf a}
\put(-150,235){\bf b}
\caption{({\bf a}) Histogram of the kink pair numbers with various quench rates. The solid lines are from the Gaussian approximations; ({\bf b}) Double logarithmic plot of the first three cumulants against the quench rate. They are almost parallel to the reference line (dashed line) with the slope $-0.25$. The solid lines are the fitted lines. }\label{k123}
\end{figure}

Fig.\ref{k123}(a) exhibits the histogram of the probabilities of the kink pair numbers with various quench rates. The solid lines are from the Gaussian approximations in the Eq.\eqref{gau} in the main text. Fig.\ref{k123}(b) shows the relations between the cumulants of the kink pair numbers against the quench rate. The dashed line is the reference line with the slope $-0.25$ which is the theoretical power from GL model. In this double logarithmic plot, we see that the first three cumulants $\kappa_{1,2,3}$ are closely parallel to the reference line. Thus, we can infer that they satisfy a universal power-law scaling to the quench rate, consistent with the KZM and the predictions from \cite{delCampo:2018hpn}. The fitted lines are approximated $\kappa_1\approx 12.930\times\tau_Q^{-0.251}, \kappa_2\approx3.805\times\tau_Q^{-0.251}$ and $\kappa_3\approx0.297\times\tau_Q^{-0.252}$. Thus, the ratios between them are $\kappa_2/\kappa_1\approx0.294$ and $\kappa_3/\kappa_1\approx 0.023$, which are very close to the theoretical predictions in the Eq.\eqref{k2k1} and Eq.\eqref{k3k1} in the main text. This indicates that the distributions of the kink numbers satisfy the binomial distributions. 

\begin{figure}[h]
\centering
\includegraphics[trim=3.2cm 9.3cm 3.4cm 9.5cm, clip=true, scale=0.55]{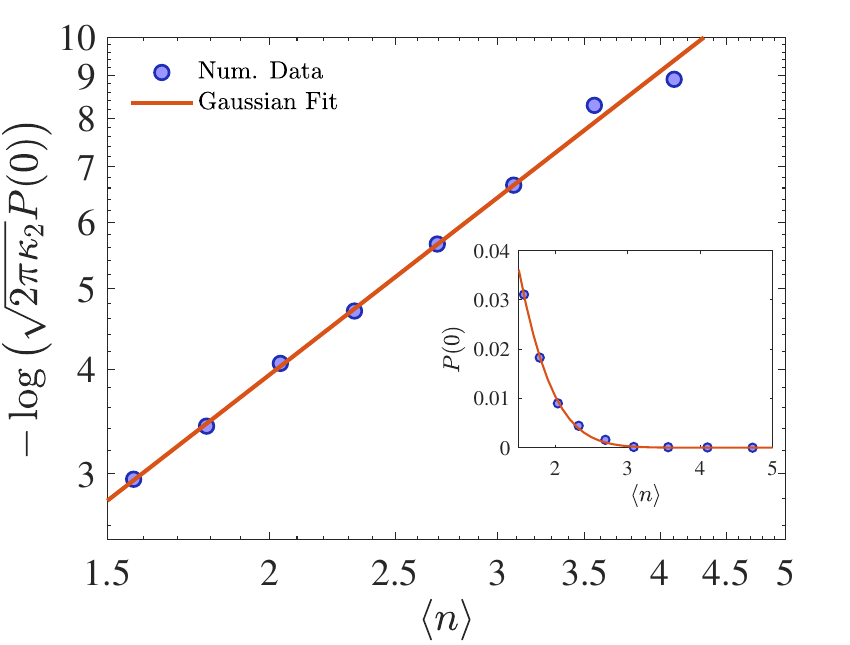}
\caption{Double logarithmic plot of the relation between $-\log\left(\sqrt{2\pi\kappa_2}P(n=0)\right)$ and the mean kink number $\langle n\rangle$, which satisfy the Gaussian approximation. The inset plot is the relation between $P(0)$ and $\langle n\rangle$ in the normal coordinates.  }\label{p0}
\end{figure}

The probability of vanishing kinks $P(n=0)$ can uncover the limit of adiabaticity of the critical dynamics. Therefore, from the Gaussian approximation Eq.\eqref{gau} in the main text, $P(n=0)$ should have a Gaussian relation to the mean kink numbers. Fig.\ref{p0} shows the double logarithmic plot between $-\log\left(\sqrt{2\pi\kappa_2}P(n=0)\right)$ and $\langle n\rangle$, which verifies the relation in Eq.\eqref{gau} in the main text. The inset plot is the relation between $P(n=0)$ and $\langle n\rangle$ in the normal coordinates.

Compared to the holographic model of the kink formations in the main text, we can see that the TDGL model has many similar behaviors to the holographic model, such as the time evolutions of the order parameter, the power-law scalings between the cumulants of the kink pair numbers vs. the quench rate, and the Gaussian approximations in the limit of vanishing kinks. We think this resemblance is not surprising, since from AdS/CFT correspondence the boundary field theory is a mean field theory with strong couplings \cite{Zaanen:2015oix}. As we have mentioned in the ``Discussion" part in the main text, it seems that the strong coupling does not change the universal power-law scalings from the theory with weak couplings. The resemblance between the strongly coupled holographic models and the weakly coupled mean field theory were commonly seen in exisiting literatures, such as in \cite{Sonner:2014tca, Chesler:2014gya}.


\end{document}